\documentclass{article}
\usepackage[numbers]{natbib}
\usepackage {arxiv}

\usepackage{cite}
\usepackage{float}
\usepackage{amsmath,amssymb,amsfonts}
\usepackage{algorithmic}
\usepackage{graphicx}
\usepackage{textcomp}
\usepackage{booktabs}
\usepackage{amsmath}
\usepackage{subfigure}
\usepackage[table,xcdraw]{xcolor}
\usepackage{soul}
\usepackage{authblk}

\begin{document}

 \title{\LARGE \bf Vine Copulas for Analyzing Multivariate Conditional Dependencies in Electronic Health Records Data}
 
 \author[1]{Manar D. Samad}
\author[1]{Yina Hou}
\author[2]{Megan A. Witherow} % Author with two affiliations
\author[3]{Norou Diawara} % Author with two affiliations
% Affiliation definitions
\affil[1]{Department of Computer Science\\
Tennessee State University\\
Nashville, TN, USA}
\affil[2]{Office of Enterprise Research \& Innovation\\
Old Dominion University\\
Suffolk, VA, USA}
\affil[3]{Department of Mathematics and Statistics \\
Old Dominion University\\
Norfolk, VA, USA}

\maketitle

\begin{abstract}

Electronic health records (EHR) store hundreds of demographic and laboratory variables from large patient populations. Traditional statistical methods have limited capacity in processing mixed-type data (continuous, ordinal) and capturing non-linear relationships in large multivariate data when oversimplified assumptions are made about the distribution (e.g., Gaussian) of disparate variables in EHR data. This paper addresses the limitations mentioned above by repurposing the vine copula method, which is primarily used to synthesize a multivariate distribution from many bivariate cumulative distribution functions (copulas). Vine copulas produce tree structures that represent bivariate conditional dependencies at varying hierarchical levels, decomposing a multivariate distribution. The tree structure is used to rank variables by conditional dependence and to identify a subset of central variables with local dependence, thus simplifying probabilistic mining of high-dimensional EHR data. The proposed application of vine copulas is used to identify conditional dependence between co-morbid conditions and is validated for characterizing different cohorts of EHR patients. The contribution of this paper is a novel approach to probabilistic mining and exploration of healthcare data that provides data-driven explanations, visualization, and variable selection to prognosticate a healthcare outcome. The source code is shared publicly in \footnote{https://github.com/mdsamad001/Vine-Copula-for-Conditional-Dependency-Analysis}.

\end{abstract}


\section{Introduction}

In healthcare informatics, modeling and analyzing multivariate health records data is a critical step in stratifying and explaining patient risk. Probabilistic modeling of data is imperative for patient risk and likelihood assessments in healthcare. However, modeling a multivariate joint distribution can be statistically complex and computationally prohibitive due to the conditional dependencies between variables. For example, the popular Naive Bayes classifier in machine learning assumes simplified conditional \emph{independence} between variables to avoid computing conditional distributions in the Bayes rule. Furthermore, many biostatistical methods used in health science arbitrarily assume normally distributed data, variable independence, and a simplified linear association between variables (e.g., Pearson's correlations). These theoretical assumptions and choices of convenience limit our ability to accurately analyze complex health data, which can lead to inaccurate and misleading conclusions. 

Vine copula models in statistics address the theoretical and computational limitations of traditional probabilistic models. Vine copulas decompose complex multivariate distributions into a series of simpler bivariate and conditional bivariate distributions structured in hierarchical trees. Vines are graph-like structures constructed hierarchically to capture variable dependencies using bivariate copulas. In contrast to traditional probabilistic methods, vine copulas offer four major benefits. First, the best-fit distribution function is obtained directly from the data instead of being arbitrarily selected. Second, disparate distribution functions are transformed into cumulative distributions, enabling efficient normalization and analysis of multivariate distributions. Third, tail dependencies and non-linear associations between variables that would be missed by the standard Pearson correlation can be captured by copula functions. Fourth, mixed-type variables, including continuous and discrete (e.g., ordinal) in a multivariate setting, can be modeled using vine copulas.

\subsection {Preliminaries on copulas} \label{prelim}

Given $\boldsymbol{X} = (X_1,\dots,X_d)^\top$ as a $d$-dimensional continuous random vector with joint cumulative distribution function (cdf) $F$ and marginal cdfs $F_j(x_j) = \mathbb{P}(X_j \le x_j)$, $j = 1,\dots,d$, we define the probability integral transforms \[
U_j = F_j(X_j), \quad j = 1,\dots,d,
\] so that $U_j \sim \mathcal{U}(0,1)$ and $\boldsymbol{U} = (U_1,\dots,U_d)^\top$ lies in $[0,1]^d$. By Sklar's theorem, the joint cdf $F$ can be written as
\begin{equation}
F(x_1,\dots,x_d) = C\big(F_1(x_1),\dots,F_d(x_d)\big),
\label{eq:cdf}
\end{equation}
where $C:[0,1]^d \to [0,1]$ is a $d$-dimensional copula. If the marginals are absolutely continuous, then the joint density $f$ admits the factorization
\begin{equation}
f(x_1,\dots,x_d)
= c\big(u_1,\dots,u_d\big)\,\prod_{j=1}^d f_j(x_j),
\label{eq:sklar-density}
\end{equation}
where $f_j$ are the marginal densities, $u_j = F_j(x_j)$, and
\[
c(u_1,\dots,u_d)
= \frac{\partial^d}{\partial u_1 \cdots \partial u_d}
  C(u_1,\dots,u_d)
\]
is the copula density. As a multivariate cumulative distribution function (cdf), the copula is a joint function that captures the dependence structure between variables.

If $X_1, \dots, X_n$ are continuous, then the copula $C$ is unique. Otherwise, $C$ can be uniquely determined on a $d$-dimensional rectangle $Range(F_1) \times \dots \times Range(F_d).$ When all of the marginals are discrete-valued, the multivariate probability mass function can be obtained as
\begin{equation*}
 f(x_1,\dots,x_d) = P(X_1= x_1, \dots, X_d= x_d) 
\end{equation*}
\begin{equation*}
      =\sum_{j_1=1}^{2}{{ \dots \sum_{j_d=1}^{2}{(-1)^{j_1+\dots+j_d}C(u_{1j_1},\dots,u_{nj_d})}}} ,    
\end{equation*}
where $u_{j1}=F_{j}(x_{j})$ and $u_{j2}=F_{j}(x_{j}^{-})$. Here 
$F_{j}(x_{j}^{-})$ is the left-hand limit of $F_{j}$ at $x_{j}$,
which is equal to $F_{j}(x_{j}-1)$.
When data are mixed-type (discrete and continuous),  the joint distribution induced by a copula is obtained by applying finite differences with respect to the discrete margins and partial derivatives with respect to the continuous margins.

\subsection{Contributions}

% The paper demonstrates one of the first applications of vine copulas to analyze complex relationships between variables in real-world electronic health records (EHR).

In contrast to the conventional use of vine copulas for synthetic data generation, this paper demonstrates a novel application to uncover conditional dependencies between variables in real-world electronic health records (EHR). Multivariate (high-dimensional) copulas are decomposed into bivariate tree structures to demonstrate conditional relationships between clinical variables and characterize different patient cohorts. The three key contributions are: 1) ranking the order of variables based on the conditional dependence of other variables to facilitate variable selection, 2) identifying local clusters of variables in vines with conditional dependence for disease-specific and patient cohort-specific study, and 3) analyzing co-morbidity using the conditional dependence between co-morbid conditions.

\begin{figure}[t]
\centering
\includegraphics[width=\linewidth]{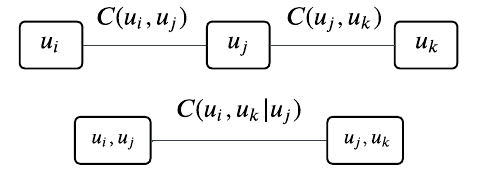}
\vspace{-10pt}
\caption{Simplified tree structures of a vine copula. Top: first tree level and bottom: second tree level. Node represents variables and edges represent bivariate copulas between nodes.}
\label{fig:copula_tree}
\vspace{-15pt}
\end{figure}

\section{Related work}

Copulas model complex multivariate data using a collection of bivariate distributions that preserve statistical dependencies among variables. Despite their popularity in statistics, few translational efforts extend the application of copulas to healthcare informatics. Bouezmarni et al. \citep{bouezmarni2025copula} have proposed a copula-based estimator to measure health inequality based on statistical dependencies between health-related variables. Copula estimators are used to determine the Gini health coefficient as a measure of inequality. In another study, an arbitrarily selected copula function (Frank copula \citep{frank1979simultaneous}) has been used to analyze the dependence between three disease outcomes (diabetes, hypertension, and osteoarthritis) in EHR data~\citep{black2018prognostic}. The dependence between disease outcomes is analyzed using the Frank distribution parameter ($\theta$) to determine whether disease pairs co-occur. The method does not account for the best-fit distribution and instead uses disease-specific univariate models for prediction. Compared with traditional multivariate copulas, vine copulas have advantages in modeling tail dependencies without distributional assumptions, modeling both continuous and ordinal variables, and capturing non-linear relationships between variables~\citep{czado2022vine}. However, vine copulas have a limited presence in healthcare informatics research. Among a few examples, R-vine copulas have been used to generate synthetic data for a randomized controlled trial (RCT) setting~\citep{petrakos2025framework}. The R-vine copula is reported to be more effective in preserving the original data distribution than sophisticated machine learning methods, such as Conditional Tabular Generative Adversarial Network (CTGAN)~\citep{xu2019modeling} and Adversarial Random Forest (ARF)~\citep{watson2023adversarial}. In healthcare, to protect patient privacy in the use of EHR, Chu et al.~\citep{chu2022vine} have used a vine copula-based perturbation in synthetic data that preserves marginal distributions and complex multivariate dependencies between discrete and continuous variables. In addition to data generation, vine copulas have also been used to stratify patient risk in healthcare. A recent study~\citep{csahin2025probabilistic} reports the first vine copula-based classifiers for probability-based patient risk profiling of perioperative patients who underwent bowel surgeries. Each of the three risk outcome classes (low, medium, high) is fitted using a separate vine copula model, which identifies low-risk patients eligible for early discharge. Although vine copulas are widely used for data generation, their potential to assess variable importance and data interpretability remains unexplored. To our knowledge, this paper presents the first application of vine copulas to the analysis of variable importance from their dependency structures in EHR data. Conditional structures in vine trees and hierarchical representations of bivariate copulas can provide useful data-driven insights into patient cohorts and disease-specific phenotypes, which may facilitate variable selection, better prognostication, efficient and more systematic analysis of large health records compared to existing statistical tools.

\section {Methodology}

This section provides a theoretical formulation of vine copulas and a methodological framework for conducting a data-driven analysis of EHR data using vine copulas.

\begin{figure}[t]
    \centering
     \subfigure[D-vine]{\includegraphics[width=0.3\textwidth]{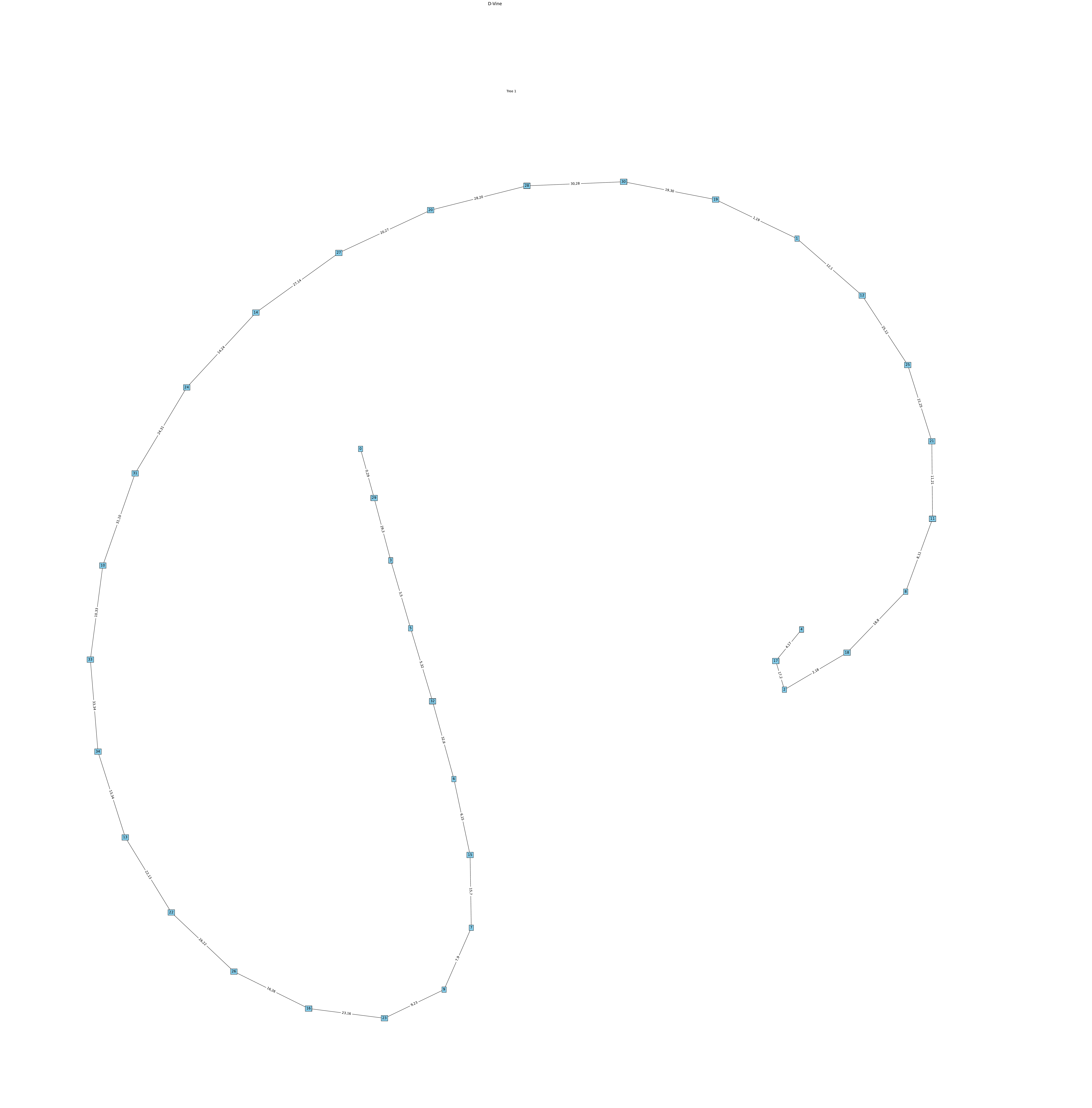}}
         \subfigure[C-vine]{\includegraphics[width=0.3\textwidth]{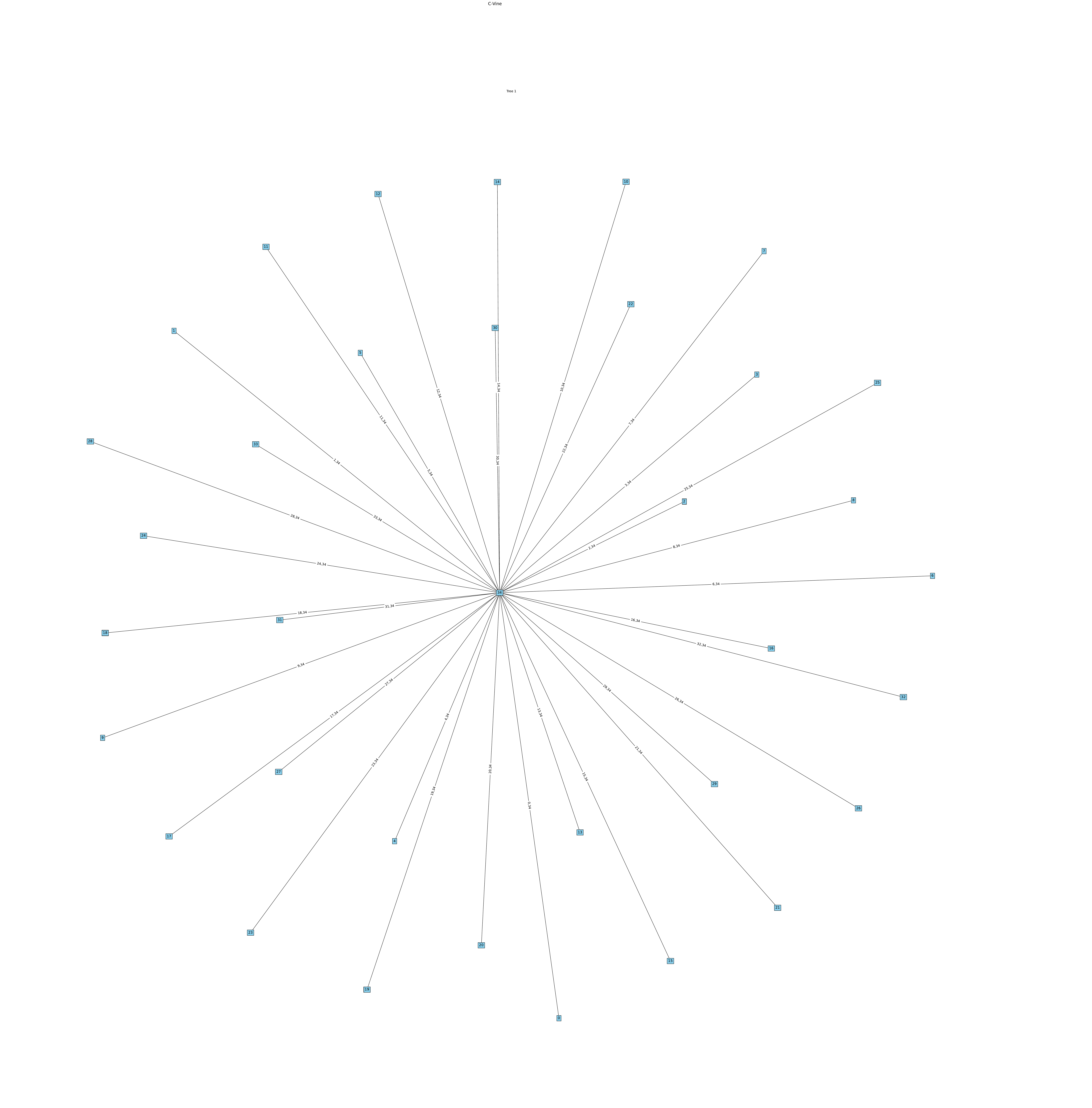}}
     \subfigure[R-vine]{\includegraphics[width=0.3\textwidth]{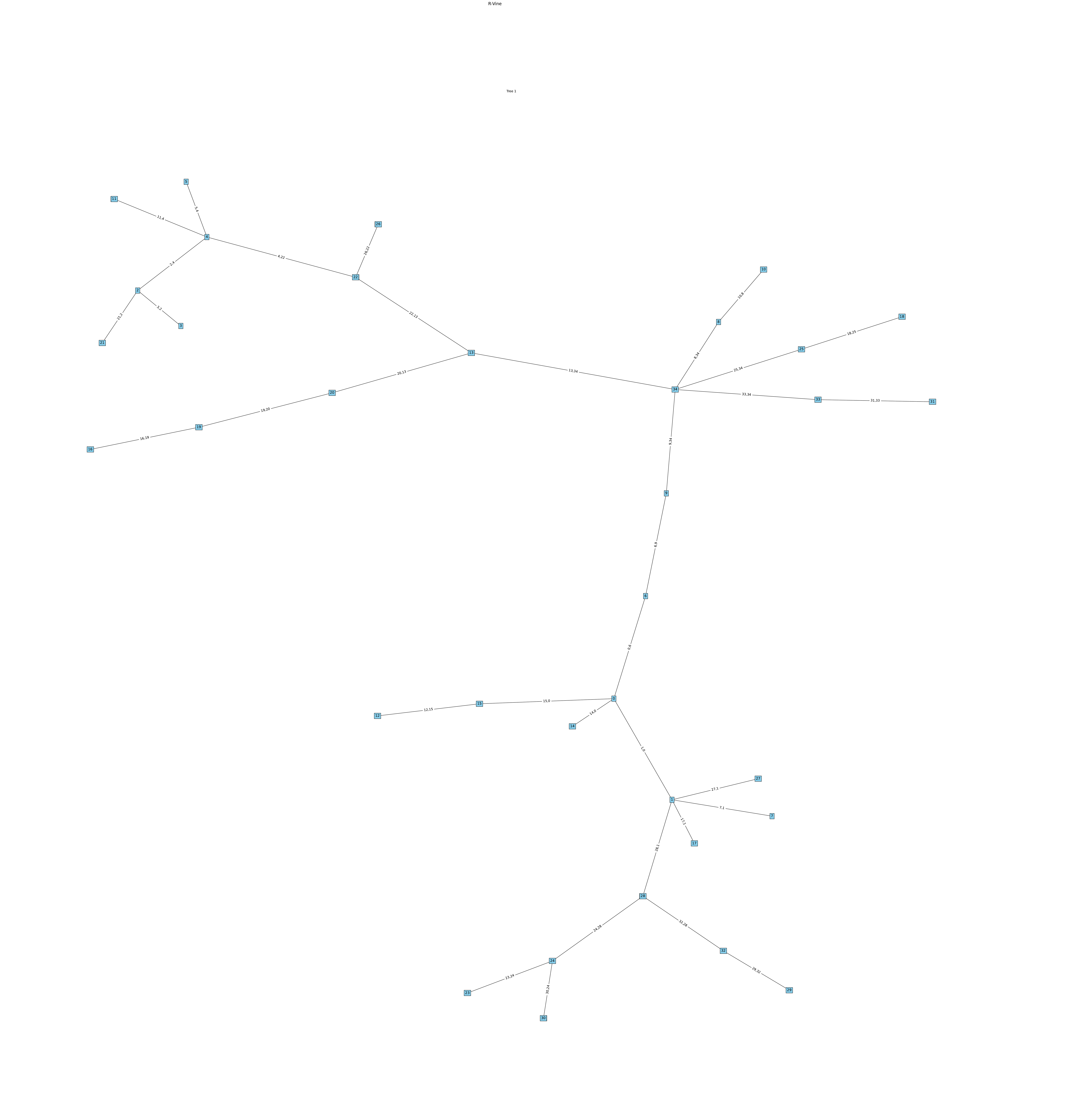}}
          \caption{The first-level tree of three types of vine copula structures. }
    \label{fig:vines}
\end{figure}

\begin{figure*}[t]
\centering
\includegraphics[width=\linewidth]{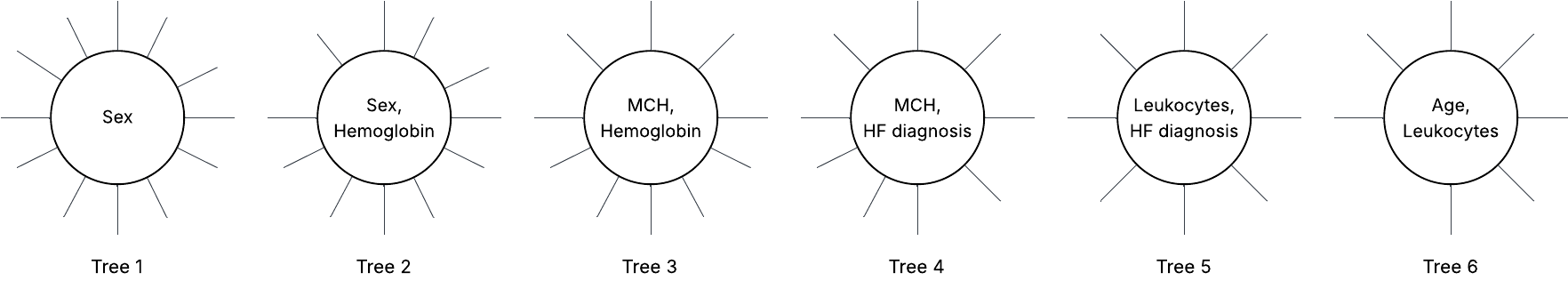}
\vspace{-5pt}
\caption{The central variables of the first six trees of C-vines are obtained from the combined patient cohort (HF, non-CVD).  The first tree shows that all other variables depend on Sex, which is the most impactful variable. MCH: Mean Corpuscular Hemoglobin. }
\label{fig:C-vines_combined}
%\vspace{-15pt}
\end{figure*}

\subsection{Vine copula formulation}

Vine copula models decompose a 
$d$-dimensional copula (Equation 1) into a sequence of bivariate pair-copulas arranged in 
a linked set of trees known as a \emph{vine}. The first-level tree structure in Figure~\ref{fig:copula_tree} shows pairwise dependencies between variables $u_i$, $u_j$, and $u_k$ (nodes) with edges that denote bivariate functions C($u_i,u_j$) and C($u_j,u_k$) as bivariate copulas. The second-level tree has nodes that denote joint bivariate dependencies ($u_i, u_j$) (edges of the first tree) and edges that denote bivariate copula C($u_i, u_k | u_j$) conditioned on the variable ($u_j$) between the nodes of the first tree.  The process of combining bivariate conditional distributions continues by gradually merging pairs of neighboring nodes creating ($d$ - 1) trees for $d$ variables. Let $\mathbf{U} = (U_1,\dots,U_d)$ denote the probability integral transforms 
of the variables.  
A vine copula factorizes the copula density as
\[
c(u_1,\dots,u_d) 
= \prod_{m=1}^{d-1} \prod_{e\in E_m} 
    c_{j_e, k_e; D_e}
    \big(u_{j_e \mid D_e}, u_{k_e \mid D_e}\big),
\]
where each factor $c_{j_e,k_e;D_e}$ is a connecting bivariate copula family 
(Gaussian, Clayton, Gumbel, Frank, etc.) that is conditioned on the set
of edges, $D_e$. The best-fit bivariate copula function is obtained using the AIC or BIC criterion. The most strongly associated variable-pairs (aka node-pairs in vine trees) are identified so that the sum of Kendall's $\tau$ of a spanning tree in the Dissmann's algorithm is maximized~\citep{dissmann2013selecting}. While Pearson's correlation assumes a Gaussian distribution and is meant for continuous variables only, Kendall's $\tau$ is a non-parametric measure of correlation that does not require a distribution assumption and can be used on both continuous and ordinal variables. Vine copulas offer four advantages over basic copula models and other statistical methods. First, vine copulas support flexible dependence modeling by capturing nonlinear, non-Gaussian, and tail-dependent interactions. Second, the edges of a tree represent a single bivariate copula with parameters that describe the strength and link between the variables. Third, vine copulas can efficiently model high-dimensional multivariate data. Fourth, the best copula function is selected edge-by-edge, resulting in a data-adaptive dependence structure.
 
\subsection{Interpretation of vine copula structures} 
 
Vine copula structures can be formulated into three types: D-vine, C-vine, and R-vine. The first level trees of these structures are illustrated in Figure~\ref{fig:vines}. The D-vine or drawable vine represents a single path-connected structure useful for modeling time-series or sequential data. The C-vine has a central node with a star-like shape, where the conditional dependence is modeled around the central variable or node. The central node represents the most influential variable ($x_c$) on which all other variables are dependent as P ($x_i | x_c$), where \{$i$ = 1,..., $d$ and $i$ $\neq$ $c$\}. C-vines can model many outcomes conditioned on one central variable. The R-vine integrates both C-vines and D-vines for flexible conditional copula modeling and reveals multiple central variables (nodes) on a D-vine path. These central nodes can identify multiple key variables that influence a subset of other variables. These structural interpretations of vine structures are taken into account in subsequent analyses of EHR data. In particular, we use C and R-vines to investigate conditional dependencies in clinical variables of patients with and without heart failure diagnoses in the EHR.

\begin{table}[t]
\centering
\caption{Summary of clinical variables extracted from the All of Us patient cohort. The bound is used to determine outlier values.}
\label{tab:summary_features}
%\scalebox{0.78}{
\begin{tabular}{lllll}
\toprule
\textbf{Feature Name} & \textbf{Abbr.} & \textbf{Unit} & \textbf{Range} & \textbf{Bound} \\
\midrule 
Age & Age & Y & 18 - 87 & 18 - 110 \\
Chloride & Cl & mmol/L & 81 - 120 & 80 - 130 \\
Sodium & Na & mmol/L & 110 - 153 & 110 - 160 \\
Calcium & Ca & mg/dL & 5.1 - 12.8 & 4 - 15 \\
Potassium & K & mmol/L & 2.1 - 10 & 2 - 10 \\
Urea nitrogen & BUN & mg/dL & 2 - 138 & 1 - 200 \\
Glucose & Glu & mg/dL & 30 - 842 & 20 - 1250 \\
Height & Height & cm & 121.92 - 213.36 & 120 - 230 \\
Creatinine & Cr & mg/dL & 0.3 - 19.9 & 0.1 - 20 \\
Weight & Weight & kg & 30.12 - 327.2 & 25 - 400 \\
Protein & TP & g/dL & 3.6 - 10.9 & 3 - 12 \\
Diastolic blood pressure & DBP & mmHg & 37 - 193 & 20 - 200 \\
Hemoglobin & Hb & g/L & 45 - 192 & 30 - 250 \\
Heart rate & HR & bpm & 20 - 159 & 20 - 250 \\
Systolic blood pressure & SBP & mmHg & 64 - 241 & 50 - 300 \\
MCV & MCV & fL & 56.9 - 125.5 & 50 - 130 \\
Alkaline phosphatase & ALP & U/L & 9.4 - 1866 & 0 - 2000 \\
Aspartate aminotransferase & AST & U/L & 3 - 1965 & 0 - 2000 \\
MCH & MCH & pg & 16.2 - 40 & 15 - 40 \\
Erythrocyte distribution width & RDW & \% & 10.7 - 33.3 & 8 - 40 \\
Carbon dioxide & CO2 & mmol/L & 10 - 45 & 10 - 45 \\
Hematocrit & HCT & \% & 16.9 - 59.7 & 10 - 70 \\
Platelets & PLT\# & $10^3$/µL & 11 - 896 & 10 - 1000 \\
Leukocytes & WBC\# & $10^3$/µL & 0.76 - 83 & 0.5 - 100 \\
Alanine aminotransferase & ALT & U/L & 3 - 1873 & 0 - 2000 \\
Erythrocytes & RBC\# & $10^6$/µL & 1.75 - 7.09 & 1 - 8 \\
Respiratory rate & RR & bpm & 6 - 48 & 4 - 60 \\
Lymphocytes percent & LYMPH\% & \% & 0 - 94 & 0 - 95 \\
Basophils percent & BASO\% & \% & 0 - 5 & 0 - 10 \\
Monocytes percent & MONO\% & \% & 0 - 27.6 & 0 - 30 \\
Triglyceride & TG & mg/dL & 20 - 1677 & 20 - 2000 \\
Eosinophils percent & EOS\% & \% & 0 - 31 & 0 - 60 \\
Cholesterol HDL & HDL-C & mg/dL & 5 - 185 & 5 - 200 \\
\bottomrule
\end{tabular}%}
\vspace{-10pt}
\end{table}

\begin{figure}[t]
\centering
\includegraphics[width=0.6\textwidth]{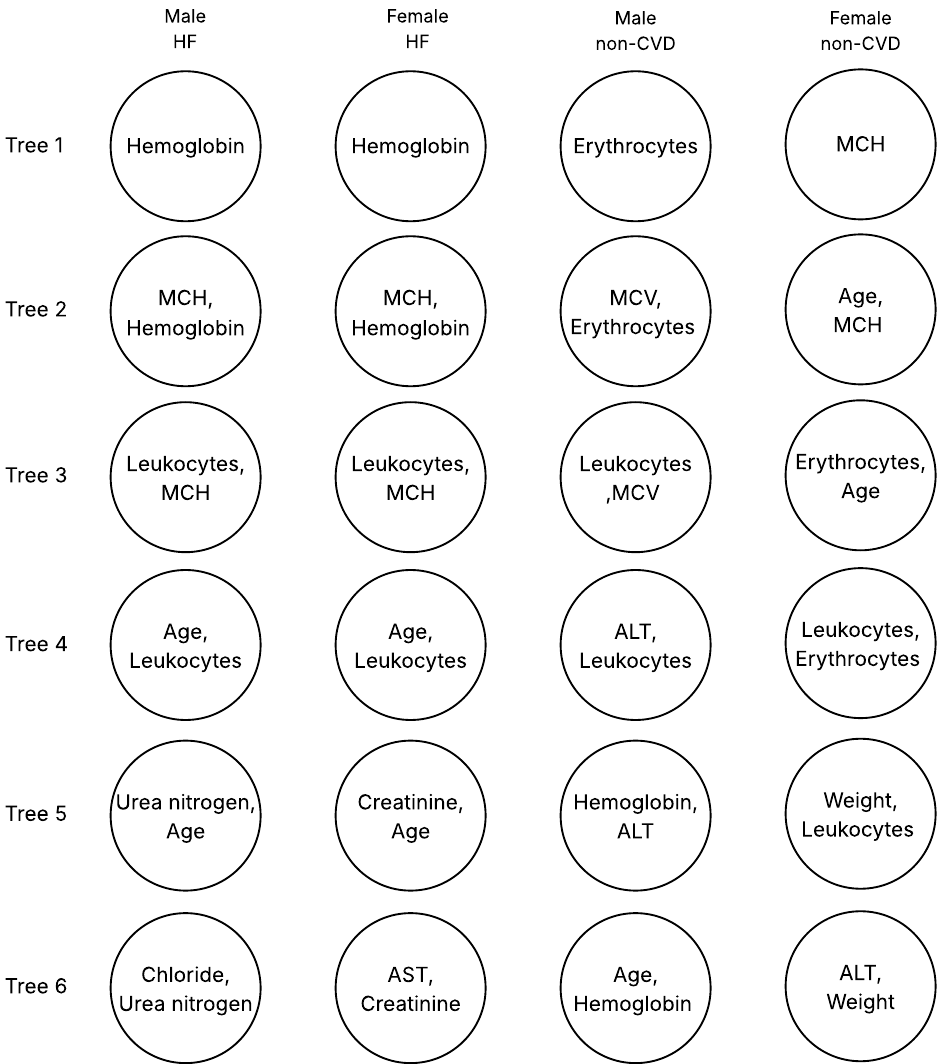}
\vspace{-5pt}
\caption{Central node of the first six C-vine bivariate copula trees conditioned on sex and heart failure diagnosis. ALT: Alanine aminotransferase, AST: Aspartate aminotransferase, MCH: Mean corpuscular hemoglobin, MCV: Mean corpuscular volume.}
\label{fig:individual_C_centers}
\vspace{-15pt}
\end{figure}

\section{EHR data extraction and curation} \label{sec:data}

EHR data used in this study are extracted and curated from the All of Us Research Program, a nationwide initiative sponsored by the US National Institutes of Health (NIH)~\citep{allofus}. The program comprises a diverse cohort of volunteers who have contributed biospecimens, physical measurements, and extensive health and lifestyle surveys to facilitate precision medicine research~\citep{sankar2017precision}. In particular, we study two different cohorts of patients: one with one or more heart failure (HF) diagnoses based on ICD-10 codes I50, I50.2-150.23, I50.3 - I50.33, I50.4 - I50.43, I50.81 - I50.814, I50.82 - I50.84, I50.8, I50.89, I50.9 and the other without ICD-10 codes (I00–I99) related to cardiovascular diseases (non-CVD group)~\citep{griffiths2004impact}. Our initial search retrieves 578,309 medical records representing 17,994 unique patients with one or more HF diagnoses. In contrast, 18,554,437 medical records in 175,530 unique patients satisfy the non-CVD patient category. Of hundreds of EHR variables, we extract 33 variables associated with HF, which are selected based on their relevance in routine healthcare and previous studies in similar patient cohorts~\citep{Luo2022, Carr2020, Zhu2023, Huang2023}. It should be noted that individual patients often have multiple records due to medical follow-ups. We use the first chronological medical record that reports an HF diagnosis. Data quality is further enhanced by applying exclusion criteria to remove outliers outside biologically plausible bounds~\citep{11153202}, as shown in Table~\ref{tab:summary_features}. In addition, patients with missing values that exceed 5\% are excluded to keep the overall missing rate below 1\% and to avoid using an external data imputation method that may compromise the data distribution. The remaining data have only 0.63\% overall missing values and are imputed using the median value of each variable. The final curated data comprises 22,679 patients or medical records, including 6,201 records with HF diagnoses and 16,478 that satisfy the non-CVD category.

\section{Results}

All access and analysis of EHR data is strictly limited to the cloud-based research workbench provided by the All of Us program. The workbench provides a cloud computing environment with a 2-core CPU with 13 GB of RAM. The vine copula models are implemented using the vinecopulas package~\citep{claassen2024vinecopulas} in the Python programming language. The \emph{best-fit} bivariate copula is identified using the AIC or BIC scores from four well-known copula functions: Gaussian, Frank, Clayton, and Student-t.

\begin{figure}[t]
%\centering
\includegraphics[width=1.02\linewidth]{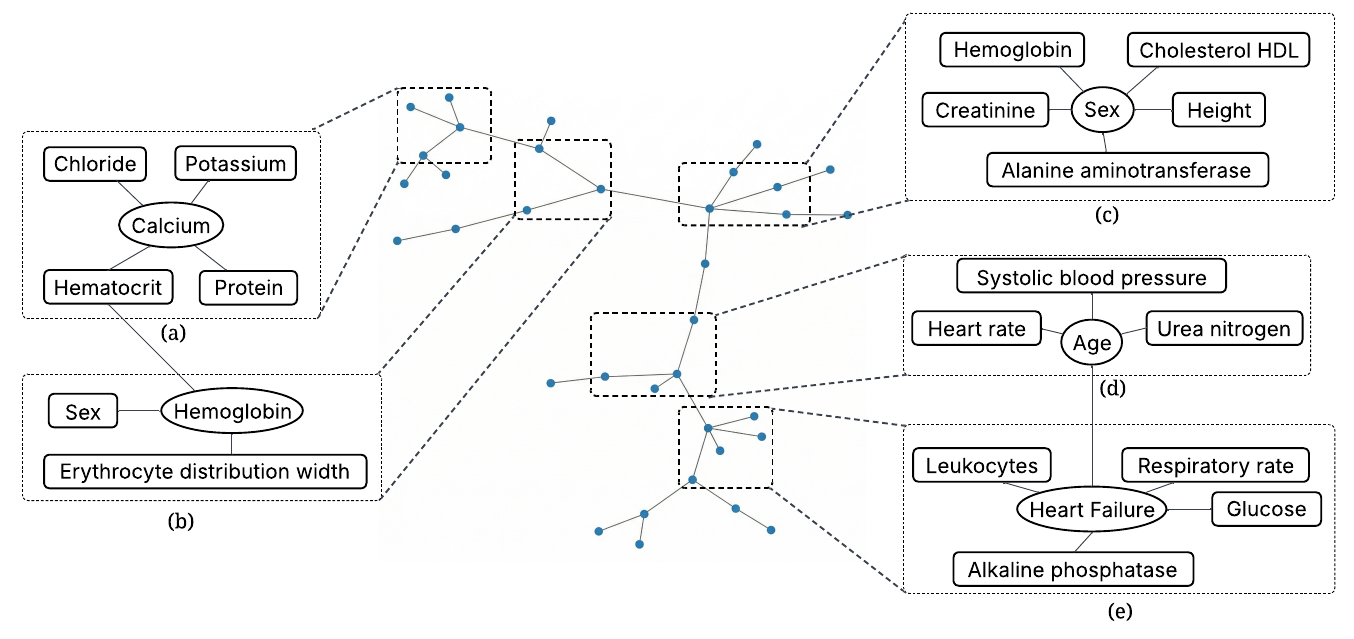}
\vspace{-20pt}
\caption{Subgraphs centered on individual variables in the first R-vine tree of combined HF and non-CVD patient cohorts.}
\label{fig:R-vine_combined}
\vspace{-15pt}
\end{figure}

\subsection {C-vine for variable selection and verification}

The C-vine structures model the conditional dependence of the remaining variables on a central variable, which influences the radial structure of the first tree. The second tree is centered on two (bivariate) variables ($x_i, x_j$), including the central variable of the first tree. The third tree will have two variables ($x_j, x_k$), and so on. The central nodes of the first six C-vine trees, obtained using a combined cohort of HF and non-CVD, are presented in Figure~\ref{fig:C-vines_combined}. The first tree shows that sex is the most central variable, affecting the distribution of all other variables. That is, sex is the most important variable that determines the values and distributions of other clinical variables. This finding is consistent with the well-known fact that the distributions of clinical variables and laboratory measurements differ between male and female patients, such that P($x_i |$ sex = male) $\neq$ P($x_i |$ sex = female). Notably, the data set includes 2.1 times more (15,346: 7,333) female patient samples than their male counterparts. We have constructed the C-vine using a 1:1 male-to-female patient ratio to confirm that sex remains the most central variable and is therefore unaffected by the imbalanced sex distribution. The second and third important variables are hemoglobin and mean corpuscular hemoglobin, which appear as the second and third most impactful variables in the subsequent trees, respectively. HF diagnosis is the fourth variable in the order. A post-hoc analysis is conducted to verify the effect of sex as a biological variable. When EHR data are grouped by sex, the accuracy of K-means clustering of HF and non-CVD patients is 75.2\% within the male cohort, and 72.2\% within the female cohort. However, when male and female cohorts are combined, including sex as a biological variable, the accuracy of HF clustering drops to 68.7\%, despite 1.5 to three times more patient samples than sex-specific cohorts~\citep{deepcluster}.  

% \begin{figure}[t]
% \centering
% \includegraphics[width=0.40\textwidth]{figure/gender_hf_C_6.pdf}
% \caption{Central node of the first six C-vine bivariate copula trees conditioned on sex and heart failure diagnosis. ALT: Alanine aminotransferase, AST: Aspartate aminotransferase, MCH: Mean corpuscular hemoglobin, MCV: Mean corpuscular volume.}
% \label{fig:individual_C_centers}
% \vspace{-15pt}
% \end{figure}

We further analyze clinical variables conditioned on sex \{male, female\} and heart failure diagnosis \{HF, non-CVD\}. Figure~\ref{fig:individual_C_centers} shows the central variables (nodes) of the first six C-vine trees conditioned on sex and diagnostic status. A consistent pattern emerges in the selection and order of variables \{Hemoglobin, MCH, Leukocytes, Age\} in both male and female patients with HF diagnosis. Clinical anemia due to reduced hemoglobin levels is observed in up to 70\% of patients with HF diagnosis~\citep{siddiqui2022anemia}. The variable after hemoglobin is MCH, which is measured as the average hemoglobin per red blood cell. Reduced MCH concentration is defined as hypochromia, which is prevalent among patients with acute HF and an independent predictor of mortality in this population~\citep{kleber2019relative}. The next variable, leukocyte count, is a marker of inflammation, which is known to be prevalent in patients with HF and predictive of adverse outcomes~\citep{zhu2021leukocyte, engström2009leukocyte}. In general, clinical measurements in HF patients, both male and female, depend more on hemoglobin, MCH, and leukocytes than they do on age. In contrast, clinical variables in females without cardiovascular disease (non-CVD) depend primarily on age after MCH. For the male cohort without CVD, clinical variables depend on erythrocytes (red blood cells), mean corpuscular volume (MCV) and leukocytes. It should be noted that patients in the non-CVD cohorts are not completely healthy, as they have ICD-10 codes for other non-CVD medical conditions.

\begin{figure}[t]
    %\centering
     \subfigure[Male with heart failure]{\includegraphics[width=0.5\textwidth]{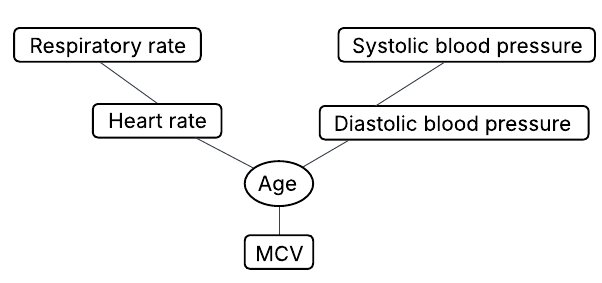}}
         \subfigure[Female with heart failure]{\includegraphics[width=0.5\textwidth]{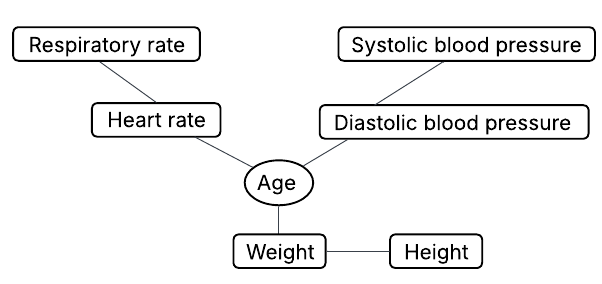}}
         \\
     \subfigure[Male without CVD]{\includegraphics[width=0.5\textwidth]{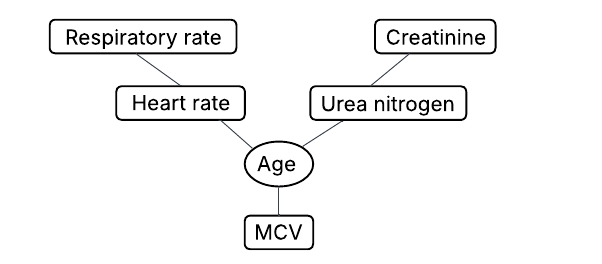}}
          \subfigure[Female without CVD]{\includegraphics[width=0.5\textwidth]{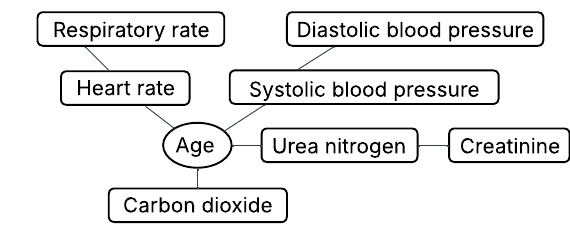}}
          \vspace{-8pt}
          \caption{Variable clusters centered on age in the first tree of the R-vine structure of four patient cohorts to show the age dependency of clinical variables. }
    \label{fig:age_tree}

\end{figure}

\subsection {R-vine for identifying conditional variable clusters}
The R-vine structures combine sequential D-vine and radial C-vine structures and are suitable for identifying clusters of conditional variables. Whereas C-vine trees capture conditional dependencies in a one-versus-all fashion, the first tree in an R-vine highlights clusters of interrelated variables. Figure~\ref{fig:R-vine_combined} shows the first tree of R-vine copulas using combined cohort data including HF and non-CVD patients. Five clusters of variables emerge centered on five variables (nodes) in the R-vine tree. The sub-figure $a$ shows several serum variables, including hematocrit and protein related to calcium, consistent with the findings that plasma protein represents the main binding substance of calcium~\citep{drop1985ionized}.

Sub-figure $b$ illustrates that hematocrit and erythrocytes are related to hemoglobin, since erythrocytes (red blood cells) are filled with hemoglobin, and hematocrit is related to hemoglobin. Creatinine, hemoglobin, and HDL cholesterol are dependent on sex (sub-figure $c$). Variables, including heart failure diagnosis, systolic blood pressure, heart rate, and urea nitrogen, depend on age (sub-figure $d$). Age and systolic blood pressure are important markers of heart failure~\citep{franklin2011aging}. Sub-figure $e$ shows the association between heart failure and age, respiratory rate. This sub-figure highlights the variables that may stratify the cohorts with and without a heart failure diagnosis. Variable groupings centered on age are further explored in different cohorts of patients based on sex \{male, female\} and diagnosis \{HF, non-CVD\}. Figure~\ref{fig:age_tree} shows the subgraphs of the first R-vine trees for four patient cohorts. 
Respiratory rate is related to heart rate, which depends on age, regardless of the cohort of the patient. MCV is exclusively dependent on age for male patients. Weight is related to age for females with a diagnosis of HF. Systolic and diastolic blood pressures show a direct dependence on age, primarily in the HF cohorts. Creatinine and nitrogen from urea are associated with age only in cohorts of patients without CVD or HF diagnosis. 
%\mds{taking examples in the previous section, see if you can find some solid clinical reference to these findings.} 

\begin{figure}[t]
    % \centering
    \subfigure[]{\includegraphics[width=0.5\textwidth]{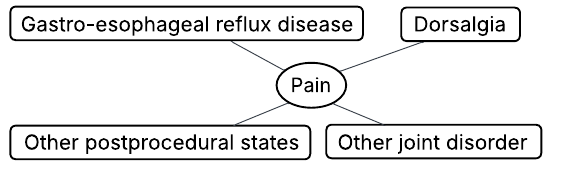}}
    \subfigure[]{\includegraphics[width=0.3\textwidth]{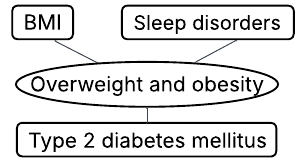}}
         % \\
    \subfigure[]{\includegraphics[width=0.5\textwidth]{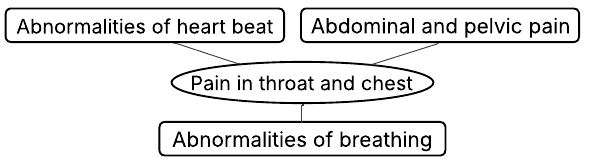}}
    \subfigure[]{\includegraphics[width=0.4\textwidth]{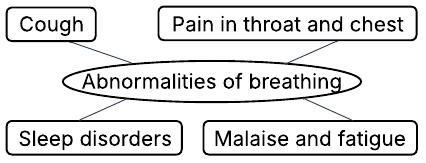}}
    \centering
    \subfigure[]{\includegraphics[width=0.5\textwidth]{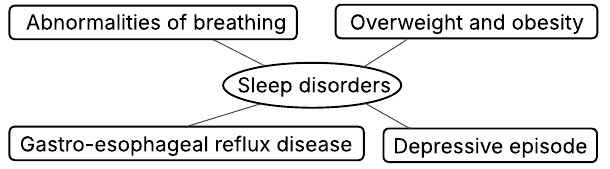}}
    \vspace{-5pt}
    \caption{Variable clusters in the first tree of R-vine structure showing co-morbidity based on ICD-10 codes.}
    %s of comorbidity analysis. Oval is the center feature for each cluster. Pain*: Pain, not elsewhere classified. Other joint disorder*: Other joint disorder, not elsewhere classified}
    \label{fig:comorbidity_subtree}
\end{figure}

\subsection{Conditional dependence in co-morbidity}

The first tree of R-vine is obtained using 20 co-morbid conditions labeled with distinct ICD-10 codes, as well as age, sex, and BMI. We identify 223,198 patient samples that have at least one confirmed diagnosis corresponding to the 20 selected ICD-10 codes. Figure~\ref{fig:comorbidity_subtree} shows the five main clusters of variables in the R-vine centered on the five most impactful co-morbid conditions. The structure centered on pain is related to dorsalgia (back pain), post-procedure complications, and joint disorders (Sub-figure $a$). Overweight and obesity are meaningfully related to BMI, as well as type-2 diabetes and sleep disorders (Sub-figure $b$). In particular, more than 90\% of people with type 2 diabetes are overweight or obese~\citep{davies2021semaglutide}. The relationship between obesity and sleep disorders is known to be bidirectional~\citep{figorilli2025obesity}. Pain in the throat and chest is related to abnormal heart rate, abdominal and pelvic pain, and abnormal breathing (Sub-figure $c$). The group centered on breathing abnormalities is associated with cough, sleep disorders, throat and chest pain, and fatigue (Sub-figure $d$). These symptoms are known to co-occur with breathing problems~\citep{jemt2022outcomes}. Sleep disorders are associated with abnormal breathing, episodes of depression, being overweight, and reflux disease (Sub-figure $e$), which is consistent with the findings that sleep disorders are a risk factor for incident depression~\citep {byrne2019sleep}.

\section{Conclusion}
This paper presents a new use of vine copulas to examine conditional dependencies in high-dimensional spaces, extending their role beyond traditional data generation. Whereas C-vines organize variables hierarchically to identify those that are most central and dependable, R-vines expose groups of nearby variables that depend on a single central variable. The relationships between variables and co-existing medical conditions uncovered by vine copulas have been found to be aligned with the medical literature. Overall, vine copulas offer a powerful framework for representing complex relationships among clinical variables by breaking down multivariate distributions into conditional bivariate components. A systematic comparison with existing baselines, such as partial correlation networks, graphical models, and sparse precision matrix estimation, would be an important direction for future work.

\section{Acknowledgments}

The research reported in this publication received support from the US National Science Foundation (NSF) award \# 2431058.

\bibliographystyle{elsarticle-num} 
\bibliography{ref,ClusterEmbed}

\end{document}